\documentclass{pasj00}

\begin{document}

\author{Mariko \textsc{Kimura} \altaffilmark{1,*},
        Keisuke \textsc{Isogai} \altaffilmark{1},
        Taichi \textsc{Kato} \altaffilmark{1},
        Akira \textsc{Imada} \altaffilmark{2},
        Naoto \textsc{Kojiguchi} \altaffilmark{3},
        Yuki \textsc{Sugiura} \altaffilmark{3},
        Daiki \textsc{Fukushima} \altaffilmark{3},
        Nao \textsc{Takeda} \altaffilmark{3},
        Katsura \textsc{Matsumoto} \altaffilmark{3},
        Shawn \textsc{Dvorak} \altaffilmark{4},
        Tonny \textsc{Vanmunster} \altaffilmark{5},
        Pavol A. \textsc{Dubovsky} \altaffilmark{6},
        Igor \textsc{Kudzej} \altaffilmark{6},
        Ian \textsc{Miller} \altaffilmark{7},
        Elena P. \textsc{Pavlenko} \altaffilmark{8},
        Julia V. \textsc{Babina} \altaffilmark{8},
        Oksana I. \textsc{Antonyuk} \altaffilmark{8},
        Aleksei V. \textsc{Baklanov} \altaffilmark{8},
        William L. \textsc{Stein} \altaffilmark{9},
        Maksim V. \textsc{Andreev} \altaffilmark{10,11},
        Tam\'as \textsc{Tordai} \altaffilmark{12},
        Hiroshi \textsc{Itoh} \altaffilmark{13},
        Roger D. \textsc{Pickard} \altaffilmark{14,15},
        Daisaku \textsc{Nogami} \altaffilmark{1}
        }
\email{mkimura@kusastro.kyoto-u.ac.jp}

\altaffiltext{1}{Department of Astronomy, Graduate School 
of Science, Kyoto University, Oiwakecho, Kitashirakawa, 
Sakyo-ku, Kyoto 606-8502}

\altaffiltext{2}{Kwasan and Hida Observatories, Kyoto 
University, Yamashina, Kyoto 607-8471}

\altaffiltext{3}{Osaka Kyoiku University, 4-698-1 Asahigaoka, 
Kashiwara, Osaka 582-8582, Japan}

\altaffiltext{4}{Rolling Hills Observatory, 1643 
Nightfall Drive, Clermont, Florida 34711, USA}

\altaffiltext{5}{Center for Backyard Astrophysics (Belgium), 
Walhostraat 1A, B-3401, Landen, Belgium}

\altaffiltext{6}{Vihorlat Observatory, Mierova 4, Humenne, 
Slovakia}

\altaffiltext{7}{Furzehill House, Ilston, Swansea, 
SA2 7LE, UK}

\altaffiltext{8}{Crimean Astrophysical Observatory, 
298409, Nauchny, Republic of Crimea}

\altaffiltext{9}{Center for Backyard Astrophysics, 6025 Calle 
Paraiso, Las Cruces, New Mexico 88012, USA}

\altaffiltext{10}{Institute of Astronomy, Russian Academy of 
Sciences, 361605 Peak Terskol, Kabardino- Balkaria, Russia}

\altaffiltext{11}{International Center for Astronomical, 
Medical and Ecological Research of National Academy of 
Sciences of Ukraine (NASU), 27 Akademika Zabolotnoho Street, 
03680 Kiev, Ukraine}

\altaffiltext{12}{Polaris Observatory, Hungarian Astronomical 
Association, Laborc utca 2/c, 1037 Budapest, Hungary}

\altaffiltext{13}{Variable Star Observers League in Japan (VSOLJ), 
1001-105 Nishiterakata, Hachioji, Tokyo 192-0153}

\altaffiltext{14}{The British Astronomical Association, 
Variable Star Section (BAA VSS), Burlington House, Piccadilly, 
London, W1J 0DU, UK}

\altaffiltext{15}{3 The Birches, Shobdon, Leominster, Herefordshire, 
HR6 9NG, UK}

\title{ASASSN-15jd: WZ Sge-type star with intermediate superoutburst between 
single and double ones}

\Received{} \Accepted{}


\SetRunningHead{Kimura et al.}{The 2015 superoutburst of ASASSN-15jd}

\maketitle

\begin{abstract}

   We present optical photometry of a WZ Sge-type dwarf nova 
(DN), ASASSN-15jd.  Its light curve showed a small dip in the middle 
of the superoutburst in 2015 for the first time among WZ Sge-type DNe.  
The unusual light curve implies a delay in the growth of the 
3:1 resonance tidal instability.  Also, the light curve is similar to 
those of other two WZ Sge-type stars, SSS J122221.7$-$311523 and OT 
J184228.1$+$483742, which are believed to be the best candidates for 
a period bouncer on the basis of their small values of the mass ratio 
($q \equiv M_{2}/M_{1}$).  
Additionally, the small mean superhump amplitude ($<$ 0.1 mag) and the 
long duration of no ordinary superhumps at the early stage of the 
superoutburst are common to the best candidates for a period bouncer.  
The average superhump period was $P_{\rm sh}$ = 0.0649810(78) d and 
no early superhumps were detected.  
   Although we could not estimate the mass ratio of ASASSN-15jd with 
high accuracy, this object is expected to be a candidate for a 
period bouncer, a binary accounting for the missing population of 
post-period minimum cataclysmic variables, based on the above 
characteristics.  

\end{abstract}

\section{Introduction}

   Cataclysmic variables (CVs) are close binary systems composed 
of a white dwarf (a primary star) and a typical late-type main 
sequence star (a secondary star).  The secondary star fills its 
Roche lobe and its matter flows toward the primary star via the 
Roche-lobe overflow.  The matter forms an accretion disk around 
the primary star and accretes on it through the disk.  

   Dwarf novae (DNe) are a subclass of CVs and have outbursts of 
typically 2--5 mag.  Their outbursts continue for days or weeks.  
The interval between outbursts are from several days to tens of 
years.  The outbursts are understood as a sudden release of 
gravitational energy by a sudden increase of the mass accretion 
rate, which is caused by a thermal instability in the disk 
(see \cite{war95book} for a review).  

\begin{figure}[htb]
\begin{center}
\FigureFile(70mm, 80mm){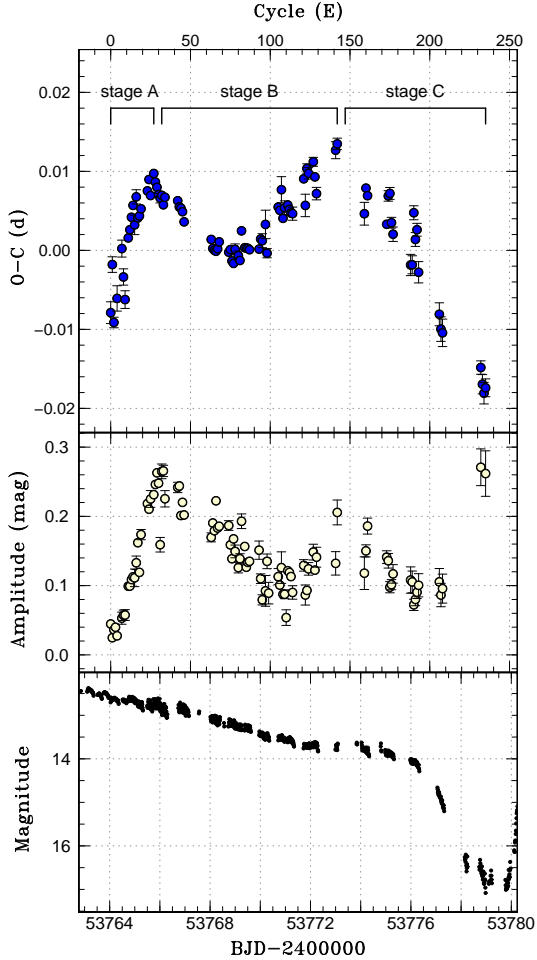}
\vspace{1cm}
\end{center}
\caption{\textcolor{black}{Exammple of the classification of superhumps by the 
variations of the period and amplitude with the actual observational 
data of the 2006 outburst in ASAS J102522$-$1542.4, a WZ Sge-type star, 
derived from Fig.~24 of \citet{Pdot3}.  Upper panel: $O - C$ curve of the 
times of superhump maxima.  
Middle panel: amplitude of superhumps.  Lower panel: light curve. 
The horizontal axis in units of BJD and cycle number of superhumps is common 
to these three panels.}}
\label{superhump}
\end{figure}

   SU UMa-type DNe with short orbital periods ($\sim$1 hr $<$ 
$P_{\rm orb}$ $<$ $\sim$ 2 hr) show 
occasional superoutbursts which are defined as outbursts with 
superhumps.  
The superhumps are believed to appear because of the 3:1 resonance 
tidal instability \citep{osa89suuma,whi88tidal,hir90SHexcess,lub91SHa,lub91SHb}.
\citet{Pdot} proposed that \textcolor{black}{the superhumps are divided 
into three stages by the variations of the period and amplitude (see 
Figure \ref{superhump} on the classification of superhumps)}.  

\begin{figure}[htb]
\begin{center}
\FigureFile(55mm, 150mm){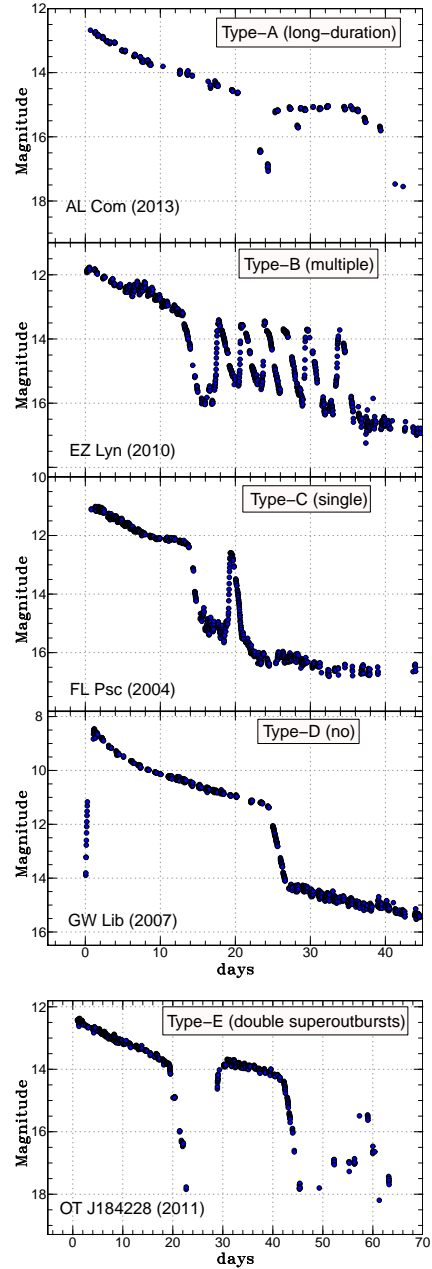}
\vspace{1cm}
\end{center}
\caption{\textcolor{black}{Classification of rebrightenings 
by their morphology.  We use the observational data in Fig.~6, 7, 
8 and 9 of \citet{kat15wzsge}.  The horizontal axis represents days 
from the starting date of their outbursts.}}
\label{rebrightening}
\end{figure}

   WZ Sge-type DNe are an extreme subclass of \textcolor{black}{SU UMa-type} 
DNe, which predominantly show superoutbursts \textcolor{black}{(for a review, 
\cite{kat15wzsge})}.  There are two important 
observational properties of them.  One is the presence of double-peaked 
modulations called early superhumps having a period almost equal to the 
orbital one in the early stage of the superoutburst 
\citep{kat02wzsgeESH,ish02wzsgeletter}.  
They are considered to be triggered by the 2:1 resonance tidal instability 
\citep{osa02wzsgehump,osa03DNoutburst}.  
The other is rebrightening after the end of the plateau stage 
\textcolor{black}{of the superoutburst}.  They 
classified the rebrightenings into five types according to the profiles 
of the light curves \citep{ima06tss0222,Pdot,Pdot5} \textcolor{black}{(see 
Figure \ref{rebrightening} on the classification of rebrightenings)}.  
The origin of the unique light curves is still an open question.  

   The evolutionary status of WZ Sge-type DNe is still in debate 
(e.g. \cite{nak14j0754j2304}).
According to the standard scenario of the CV evolution, CVs evolve as 
the orbital period becomes shorter because of angular momentum losses 
by magnetic breaking and gravitational radiation.  
Once $P_{\rm orb}$ reaches the period minimum, the secondary star 
becomes degenerate and the thermal timescale of the secondary 
star becomes longer than the mass-transfer timescale.  Hence, the 
systems evolve back toward longer orbital periods.  They are called 
period bouncers (see \cite{kni11CVdonor} and references therein).  
\citet{kol93CVpopulation} suggested that $\sim$70\% of all CVs would be 
beyond the period minimum.  It is however difficult to discover them
because they are faint and have a long recurrence time between outbursts 
\citep{pat11CVdistance}.  

   In the study of the missing population of post-period minimum CVs, 
it is worth noting that \citet{lit06j1035} demonstrated with their 
spectroscopy that the donor star in the eclipsing CV SDSS 
J103533.03$+$055158.4 with a period close to the minimum one was a 
brown dwarf, suggesting that this object is a period bouncer.  
Moreover, \citet{lit08eclCV} found three more eclipsing CVs 
possess brown dwarfs as the secondary star using high-speed, 
three-color photometry.  
   Recently, other several good candidates for a period bouncer have 
been discovered among WZ Sge-type DNe with \textcolor{black}{type-B 
rebrightening or type-E rebrightening or slow fading rate (the 
objects and references are summarized in Table 1)} 
through optical photometry using the new method of estimating the mass 
ratio ($q$) of the secondary to the primary with the period of stage A 
superhumps and the orbital period \citep{kat13qfromstageA}.  
These period-bouncer candidates have some common properties: 
(1) long-lasted stage A superhumps, (2) a large decrease of 
the superhump period between stage A and stage B, (3) a small superhump 
amplitude ($\lesssim$ 0.1 mag) and (4) long-lasting early 
superhumps ($\gtrsim$ 10 d).  
\citet{kat15wzsge} also identified the relation between the 
rebrightening types and the CV evolutionary stage (WZ Sge-type stars 
would evolve in the order of type C $\rightarrow$ D $\rightarrow$ A 
$\rightarrow$ B $\rightarrow$ E) and the two objects with type-E 
rebrightening \textcolor{black}{(see Table 1)} 
are the best candidates for a period bouncer.  

   In this paper, we report on the optical photometry of a WZ 
Sge-type object, ASASSN-15jd having similar characteristics to those of 
the objects with type-E rebrightening.  
Its outburst was detected on 2015 May 13 by All-Sky Automated 
Survey for Supernovae (ASAS-SN) \citep{ASASSN,dav15ASASSNCVAAS}.
This object has a quiescent counterpart SDSS J164925.43$+$140243.5
($g$ = 22.8 mag) and its position is (RA:) 16h49m25.42s,
(Dec:) +\timeform{14D02'43.7''} (J2000.0).  

   This paper is structured as follows.  We describe a log of observations 
and our analysis method in Section 2 and the results in Section 3.  
In Section 4, we discuss the results.  A brief summary and 
further observations are described in Section 5.  

\begin{table*}
  \caption{Properties of candidates for a period bouncer}
\label{tab:bouncer}
\begin{center}
\begin{tabular}{lllllcc}
\hline
Object$^{*}$ & $P_{\rm orb}^{\dagger}$ (d) & $P_{1}^{\ddagger}$ (d) & $P_{2}^{\S}$ (d) & Amp$^{\parallel}$ & Profile of light curve$^{\#}$ & Reference$^{\P}$\\
\hline
MASTER J1222 & 0.059732(3) & 0.06158(5) & 0.060221(9) & 0.0995 & multiple rebrightenings (type-B) & 1\\
MASTER J2037 & 0.06051(18) & 0.06271(11) & 0.061307(9) & 0.114 & multiple rebrightenings (type-B) & 1\\
SSS J1222 & 0.075879(1) & 0.07721(1) & 0.07649(1) & 0.115 & double superoutbursts (type-E) & 2\\
OT J1842 & 0.0724 & 0.07275$^{**}$ & & 0.08 & double superoutbursts (type-E) & 3\\
OT J0754 & -- & 0.072218(3) & 0.070758(6) & 0.0531 & slow decline & 4\\
OT J2304 & -- & 0.067245(17) & 0.066351(12) & 0.127 & slow decline & 4\\
ASASSN-14cv & 0.059917(4) & 0.06168(2) & 0.060450(14) & 0.0732 & multiple rebrightenings (type-B) & 5, 6\\
PNV J1714 & 0.059558(3) & 0.06130(2) & 0.060084(4) & 0.0935 & multiple rebrightenings (type-B) & 5, 6\\
OT J0600 & -- & 0.064659(12) & 0.063310(4) & 0.0648 & multiple rebrightenings (type-B) & 5, 6\\
ASASSN-15jd & -- & -- & 0.064981(8) & 0.0916 & a small dip in the middle & This paper\\
& & & & & of the plateau & \\
\hline
\multicolumn{7}{l}{
$^{*}$Objects' name; MASTER J1222, MASTER J2037, SSS J1222, OT J1842, OT J0754, OT J2304, PNV J1714} \\
\multicolumn{7}{l}{
and OT J0600 represent MASTER OT J211258.65$+$242145.4, MASTER OT J203749.39$+$552210.3,} \\
\multicolumn{7}{l}{
SSS J122221.7$-$311523, OT J184228.1$+$483742, OT J075418.7$+$381225, OT J230425.8$+$062546,} \\
\multicolumn{7}{l}{
PNV J17144255$-$2943481 and OT J060009.9$+$142615, respectively.} \\
\multicolumn{7}{l}{
$^{\dagger}$Orbital period referring to the period of early superhumps.  } \\
\multicolumn{7}{l}{
$^{\ddagger}$Period of stage A superhumps.} \\
\multicolumn{7}{l}{
$^{\S}$Period of stage B superhumps.} \\
\multicolumn{7}{l}{
$^{\parallel}$Mean amplitude of superhumps (mag).} \\
\multicolumn{7}{l}{
$^{\#}$Characteristic shapes of light curves.} \\
\multicolumn{7}{l}{
$^{\P}$1: \citet{nak13j2112j2037}, 2: \citet{kat13j1222}, 3: \citet{kat13j1842}, 4: \citet{nak14j0754j2304},} \\
\multicolumn{7}{l}{
5: Nakata et al.~in preparation, 6: \citet{Pdot7}} \\
\multicolumn{7}{l}{
$^{**}$Mean superhump period.} \\
\end{tabular}
\end{center}
\end{table*}

\section{Observation and Analysis}

   Time-resolved CCD photometry was carried out by the VSNET 
collaboration team at twelve sites (Table 2). Table 3 shows the log 
of photometric observations.  
All of the observation times were converted to barycentric 
Julian date (BJD).
Before making the analyses, we applied zero-point corrections 
to each observer by adding constants.
The magnitude scales of each site were adjusted to that of 
the Kyoto University system (KU1 in Table 3), where 
GSC974.1416 (RA: 16h49m32.51s, Dec:+\timeform{14D03'22.7''}, 
$V$ = 13.0) was used as the comparison star. 
The constancy of the comparison star was checked by nearby 
stars in the same images.  

   We used the phase dispersion minimization (PDM) method \citep{PDM}
for a period analyses.  We subtracted the global trend of the light curve 
by locally weighted polynomial regression (LOWESS: \cite{LOWESS}) 
before making the PDM analyses.  The 1$\sigma$ error of the best 
estimated period by the PDM analysis was determined by the methods 
in \citet{fer89error} and \citet{Pdot2}.  

   A variety of bootstraps was used for estimating the robustness of 
the result of PDM.  We analyzed about 100 samples which randomly 
contained 50\% of observations, and performed a PDM analysis for these 
samples.  
The result of the bootstrap is expressed as a form of 90\% confidence 
intervals in the resultant $\theta$ statistics.

\section{Results}

\subsection{Overall Light Curve}

\begin{figure}[htb]
\begin{center}
\FigureFile(80mm, 50mm){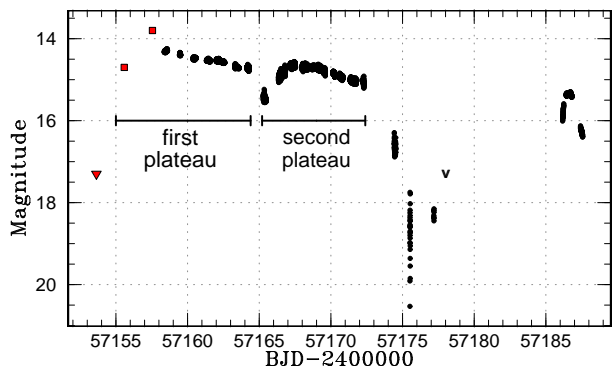}
\vspace{1cm}
\end{center}
\caption{Overall light curve of the 2015 superoutburst of ASASSN-15jd 
(BJD 2457153--2457188).  
The circles and `V'-shapes represent CCD photometric observations 
and upper limits by KU1 (ref.~Table 3), respectively.  The quadrangles 
and inverted triangles represent the detections and upper limits by 
ASAS-SN, respectively.}
\label{overall}
\end{figure}

   We show the overall light curve of the 2015 superoutburst of 
ASASSN-15jd in Figure \ref{overall}.  The superoutburst probably 
began on BJD 2457155 and the object showed a rapid rise at the very 
early stage.  
\textcolor{black}{The first plateau stage} continued for about ten days 
\textcolor{black}{during BJD 2457155--2457164.4} and a small dip whose 
depth is about 1 mag was observed on BJD 2457165.  
Immediately after the dip, the object showed a rapid rise again 
and \textcolor{black}{the second plateau stage} started.  It continued for 
about a week \textcolor{black}{during BJD 2457165.2--2457172.4} and a rapid 
fading was seen on 2457174.  There were no observations during BJD 
2457178--2457186.  On BJD 2457186, the onset of a rebrightening was 
detected (vsnet-alert 18815).  The duration of the rebrightening was 
short, a few days.  

\subsection{Superhumps}

\begin{figure}[htb]
\begin{center}
\FigureFile(80mm, 150mm){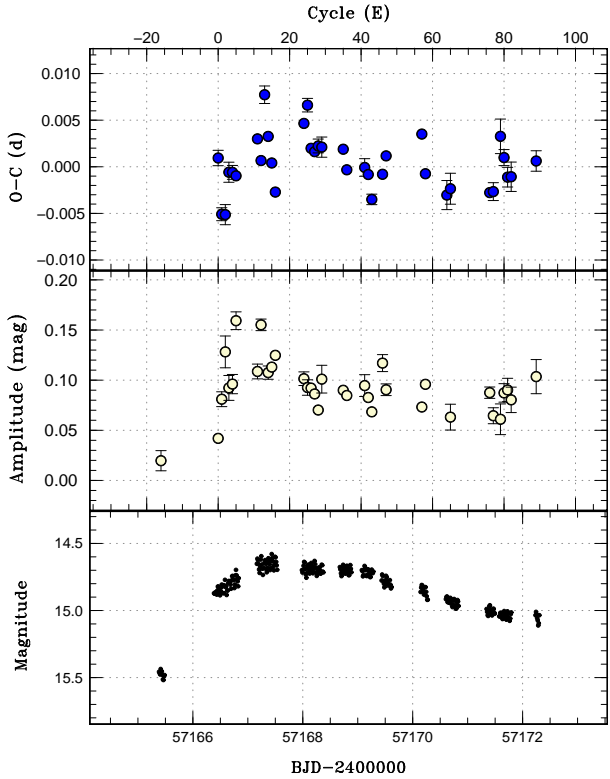}
\vspace{1cm}
\end{center}
\caption{Upper panel: $O - C$ curve of the times of superhump maxima 
of during BJD 2457165.2--2457172.4 (\textcolor{black}{the second plateau stage} 
of ASASSN-15jd).  An ephemeris of BJD 2457166.459117$+$0.0650258 E was 
used for drawing this figure. Middle panel: amplitude of superhumps.  
Lower panel: light curve. The horizontal axis in units of BJD and 
cycle number is common to these three panels.}
\label{o-c}
\end{figure}

   During the dip (on BJD 2457165), ordinary superhumps started to 
develop.  The $O - C$ curve of times of superhump maxima, the amplitude 
of superhumps and the light curve during BJD 2457165.2--2457172.4 are 
shown in the upper panel, the middle panel and the lower panel of 
Figure \ref{o-c}, respectively.  We determined the times of maxima of 
ordinary superhumps in the same way as in \citet{Pdot}.  Some points with 
large errors were removed in calculating the $O - C$ and amplitude.  
The resultant times are given in Table 4.  Although we could not clearly
identify the term of stage A only from the $O - C$ curve, we regarded 
BJD 2457166.2--2457167.6 ($0 \leq E \leq 16$) as the final part of
stage A, judging from the variations of the superhump amplitudes.  
Also, we considered that \textcolor{black}{the term of stage B was 
BJD 2457167.9--2457172.4 ($24 \leq E \leq 90$) from the nonlinear behavior 
on the $O - C$ plane and the slow decrease of superhump amplitudes.  
The stage B superhumps might continue after BJD 2457172.4 although 
we could not detect it because of the sparse coverage.}
The superhumps disappeared at the fading stage on BJD 2457174 and did 
not develop again in the rebrightening at least in our 
\textcolor{black}{low accuracy} observations and we could not find stage C 
during \textcolor{black}{the second plateau stage} and after that.  

\begin{figure}[htb]
\begin{center}
\FigureFile(80mm, 50mm){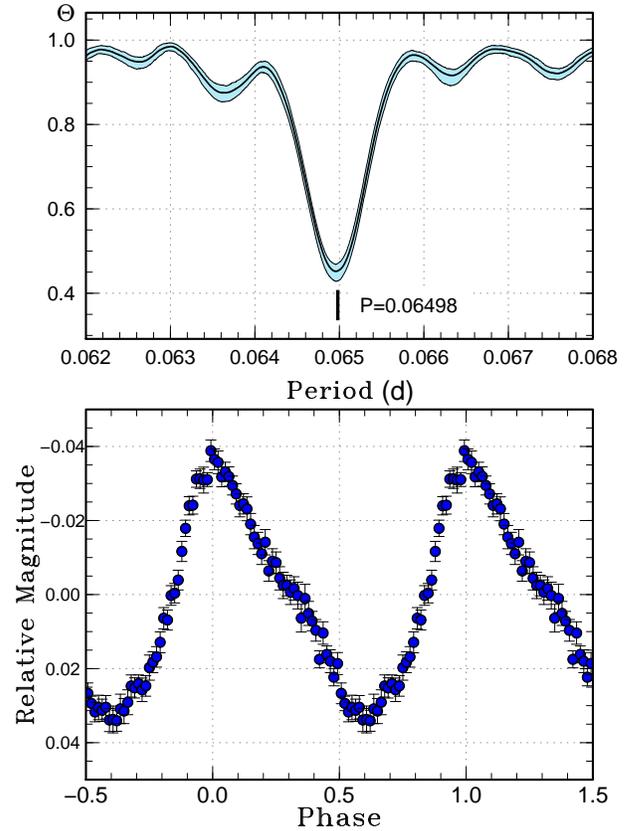}
\vspace{1cm}
\end{center}
\caption{Superhumps in \textcolor{black}{the second plateau stage} of the 2015 
outburst of ASASSN-15jd.  Upper: $\Theta$-diagram of our PDM analysis 
of stage B superhumps (BJD 2457167.9--2457172.4).  Lower: Phase-averaged 
profile of stage B superhumps.}
\label{pdm}
\end{figure}

   We applied a period analysis by the PDM method for 
stage B (BJD 2457167.9--2457172.4) and obtained a period of $P_{\rm sh}$ 
= 0.0649810(78) d (see upper panel of Figure \ref{pdm}).  Here,
the data having low accuracy were excluded from the light 
curve when we performed our PDM analysis.  The derivative of the 
superhump period during stage B was $P_{\rm dot} 
(\equiv \dot{P}_{\rm sh}/P_{\rm sh}) = 10.8 (3.8) 
\times 10^{-5} {\rm s}~{\rm s}^{-1}$.  
   The mean profile of stage B superhumps is also shown in the lower 
panel of Figure \ref{pdm}.  

\textcolor{black}{Early superhumps, double-wave modulations with 
a period almost the same as the orbital one usually appear prior to 
superhumps during superoutbursts of WZ Sge-type DNe 
\citep{kat02wzsgeESH,ish02wzsgeletter}.  
However, during the first plateau stage of the 2015 outburst 
of ASASSN-15jd (during BJD 2457155--2457164.4), any humps with an amplitude 
of $>$0.02 mag were not observed in the range of 97.5--99.5\% of the 
estimated superhump period in the previous paragraph.}  

\begin{figure}[htb]
\begin{center}
\FigureFile(80mm, 100mm){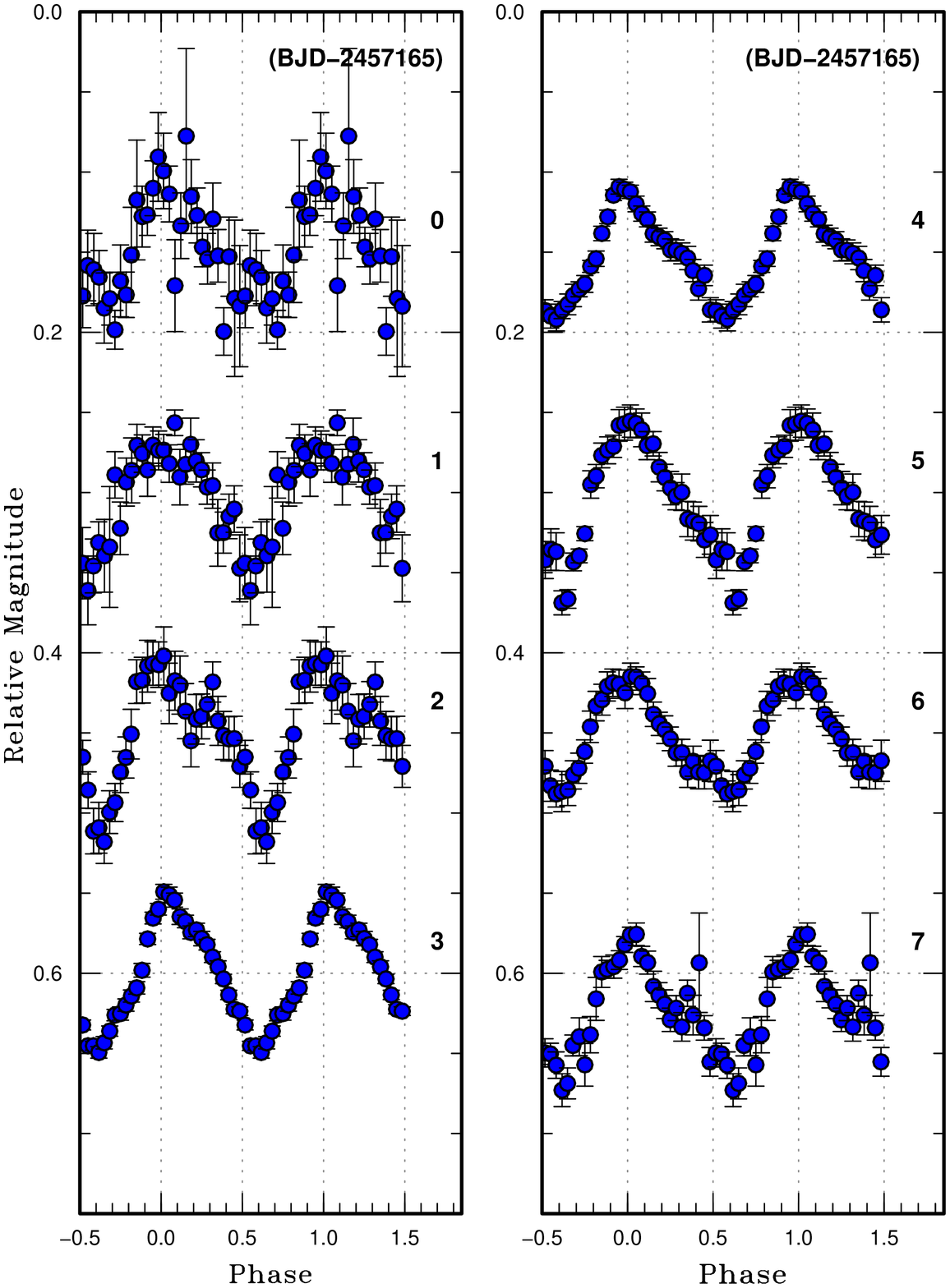}
\vspace{1cm}
\end{center}
\caption{Dairy variation of the profile of superhumps in ASASSN-15jd during 
BJD 2458165.2--2457172.4.}
\label{profile}
\end{figure}

   In Figure \ref{profile}, the daily variation of the profile of 
superhumps during BJD 2457165--2457172 is shown.  The daily 
average amplitude of superhumps was 0.08--0.12 mag.

\ifnum0=1
\subsection{Absence of Early Superhumps}

   In order to detect early superhumps, we applied the PDM analysis 
to the first half of the plateau stage at first, however, any periods 
were not detected.  Second, we assumed that the period of early superhumps 
is in the range of 97.5--99.5\% of $P_{\rm sh}$ and depicted the average 
profile of humps during BJD 2457158--2457164 for each period dividing the 
range into 200.  The mean amplitudes were, however, less than 0.02 mag.  
\citet{uem12ESHrecon} predicted that early superhumps are detected 
with an amplitude of $>$ 0.02 mag in about 90\% of WZ Sge-type DNe 
and the observed amplitudes in $\sim$60\% of the reported WZ Sge-type 
DNe are larger than 0.02 mag \citep{Pdot7}.  
Therefore, our analyses show that early superhumps were below the detection
limit in this outburst.
\fi

\section{Discussion}

\subsection{Small Dip in the Plateau of the Superoutburst}

\begin{figure*}[htb]
\begin{minipage}{0.5\hsize}
\label{typeA--D}
\begin{center}
\FigureFile(80mm, 50mm){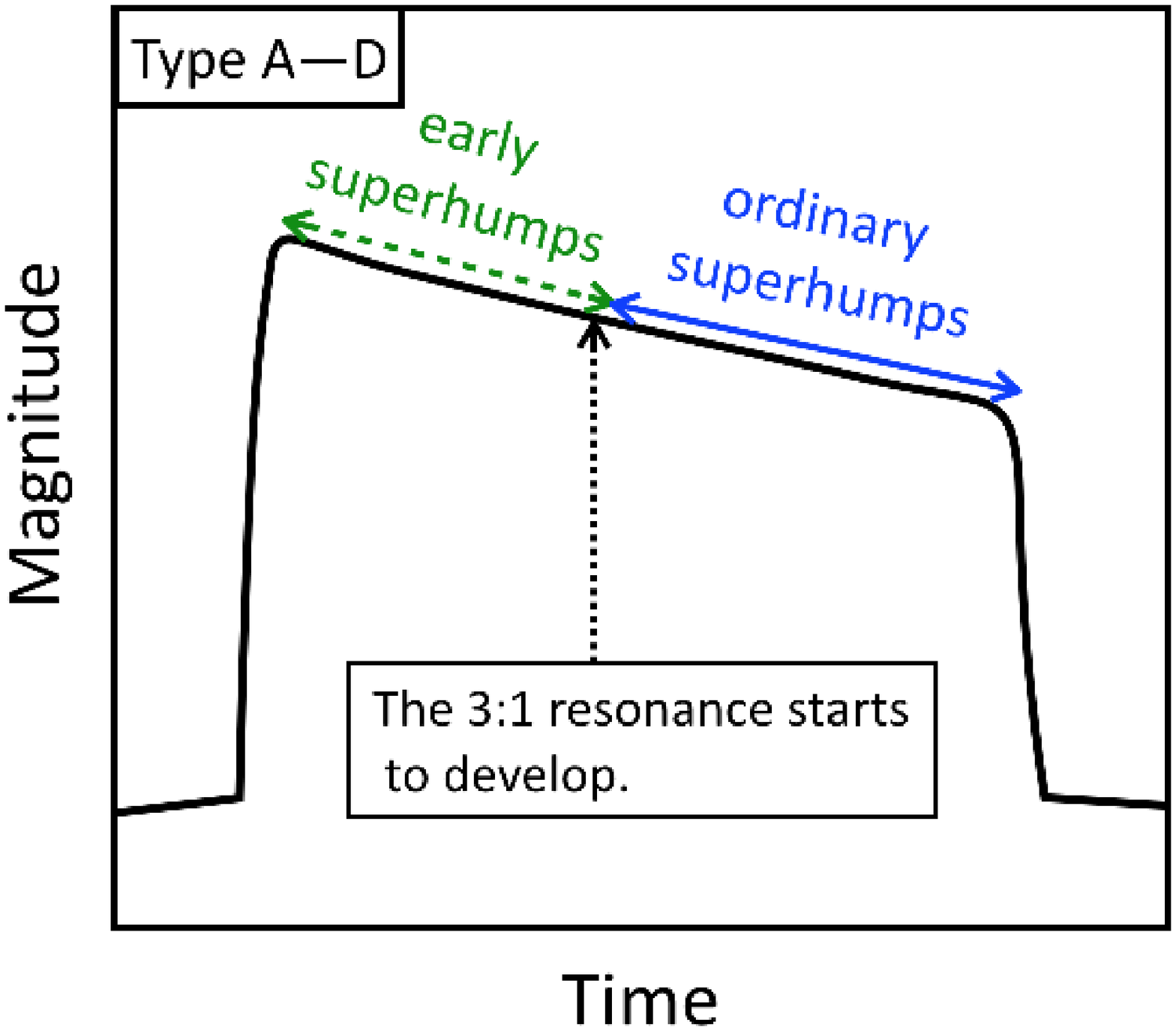}
\end{center}
\end{minipage}
\begin{minipage}{0.5\hsize}
\label{typeE}
\begin{center}
\FigureFile(80mm, 50mm){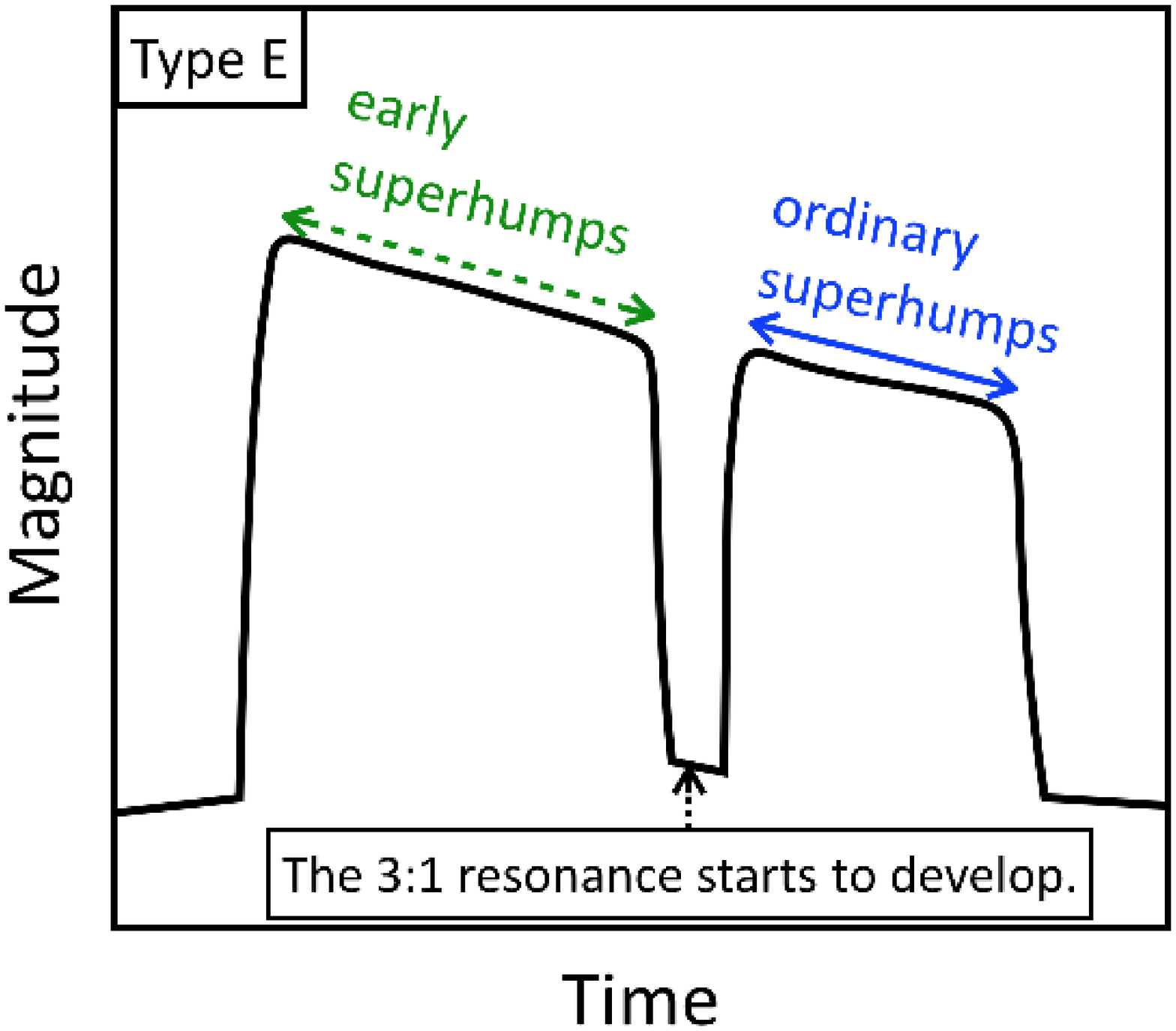}
\end{center}
\end{minipage}
\caption{Classification of plateau stages observed in WZ Sge-type DNe.  
Left: single plateau stage observed in objects with type-A, B, C 
and D rebrightenings.  Right: double plateau stages observed in 
objects with type-E rebrightening.}
\label{plateau}
\end{figure*}

   ASASSN-15jd is the first example among WZ Sge-type DNe, which 
showed a small dip of brightness in the middle of the superoutburst.  
It has been known that the plateau stage of superoutbursts in WZ Sge-type 
objects can be classified into two types: single plateau stage observed 
in the objects with type-A, B, C and D rebrightening (except for type-E 
rebrightening) and double plateau stages observed only in the objects 
with type-E rebrightening.  
The schematic diagrams of the light curves of these two plateau types 
are depicted in Figure \ref{plateau}.  The left and right panels of 
Figure \ref{plateau} represent single plateau stage and double plateau 
stages, respectively.  
In the single plateau stage, ordinary superhumps are seen as soon as 
early superhumps disappear, while in the double plateau stages, ordinary
superhumps begin to develop in \textcolor{black}{the second plateau stage} 
a few days after early superhumps disappear.  
The plateau stage in ASASSN-15jd cannot be classified as one of 
these two types of the plateau stage.  \textcolor{black}{There would be a 
possibility that the dip in ASASSN-15jd was deeper since we did not have 
adequate observations around the observed dip on BJD 2457165; neverthless, 
the duration of the dip in this object was $\sim$2 days and shorter than 
$\sim$10 days duration in the objects with type-E rebrightening 
(\cite{kat13j1222,kat13j1842} and see also the bottom panel of Figure 
\ref{rebrightening}).  Hence, we can regard that ASASSN-15jd showed an 
intermediate light curve between the single plateau stage and the double 
ones.}  

   Ordinary superhumps are considered to develop together with the 3:1 
resonance tidal instability occurring in accretion disks (see e.g.,
\cite{osa96review} for a theoretical review).  
We can therefore regard that the shapes of plateau stages in 
WZ Sge-type DNe reflect the speed at which the 3:1 resonance develops.  
In other words, it is considered that we cannot observe the dip of 
brightness in the plateau in the objects with the single plateau stage 
since the 3:1 resonance grows up as soon as the 2:1 resonance finishes
working.  
In the cases when the 3:1 resonance does not grow sufficiently quickly,
the cooling wave propagates in the accretion disk before the 3:1 resonance
fully builds up in the objects with the double plateau stages, and hence,
these have a dip of brightness.  The 3:1 resonance in ASASSN-15jd
seems to develop slower than that in the former objects (single plateau)
and faster than that in the latter objects (double plateaus).  

   There is a theory explaining the reason why the delay in 
development of superhumps appears.  The growth time of the 
3:1 resonance tidal instability is inversely proportional to the square 
of the mass ratio \citep{lub91SHa}.  The type-E rebrightening 
was observed in the outbursts of SSS J122221.7$-$311523 and 
OT J184228.1$+$483742 in the past \citep{kat13j1222,kat13j1842} and 
these systems are considered to be promising candidates for a period 
bouncer because of their small estimated mass ratios \citep{kat15wzsge}.
Their mass ratios are smaller than those of other WZ Sge-type stars with 
the single plateau, and the behavior is consistent with the slow growth
of the 3:1 resonance.  Although we could not estimate 
the accurate value of the mass ratio of ASASSN-15jd, the 3:1 
resonance seems to develop slower in this system than in WZ Sge-type 
stars with the single plateau stage judging from the shape of the 
plateau stage and it is expected that this object has a small $q$ 
value.  
Combined with the relatively long $P_{\rm sh}$ for a WZ Sge-type star, 
ASASSN-15jd is likely to be a promising candidate for a period 
bouncer.  

   The rebrightening type after the second plateau stage in this outburst 
is probably C (single rebrightening) although there remain a possibility 
of multiple rebrightenings due to the lack of observations.  This rebrightening 
type has been observed in many other WZ Sge-type objects.

\subsection{Small Amplitude of Superhumps}

   The average superhump amplitude of ASASSN-15jd in the 2015 
superoutburst was small, less than 0.1 mag (see the lower panel of 
Figure \ref{pdm} and Figure \ref{profile}).  In about 90\% of ordinary 
SU UMa-type DNe, the normalized superhump amplitudes fall within the 
range of 0.14--0.35 mag independent of the inclination \citep{Pdot3}.  

\begin{figure*}[htb]
\begin{minipage}{0.5\hsize}
\label{amp-cycle}
\begin{center}
\FigureFile(80mm, 50mm){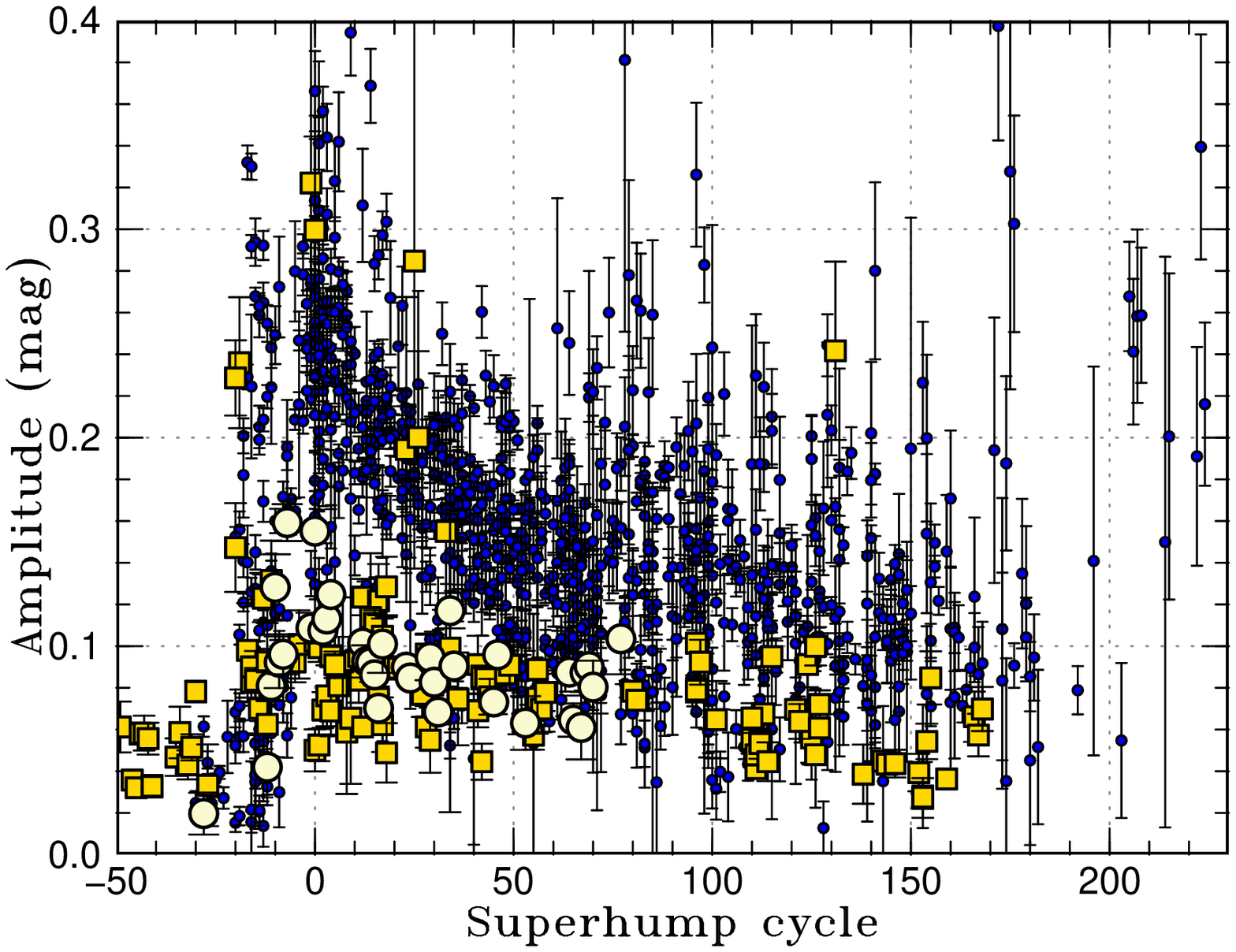}
\vspace{0.5cm}
\end{center}
\end{minipage}
\begin{minipage}{0.5\hsize}
\label{amp-porb}
\begin{center}
\FigureFile(80mm, 50mm){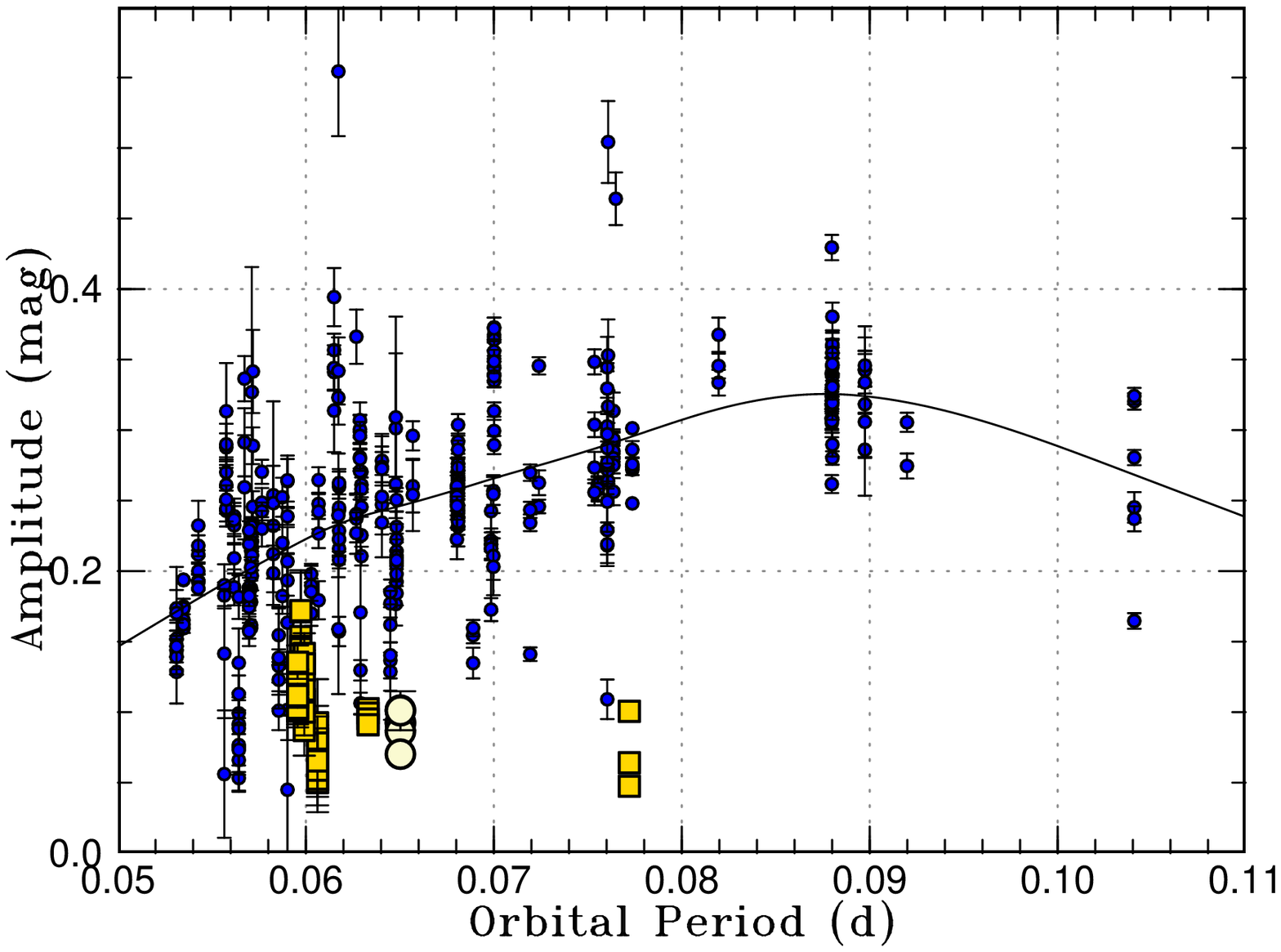}
\vspace{0.5cm}
\end{center}
\end{minipage}
\caption{Left: Variation of superhump amplitudes in the SU UMa-type 
objects with 0.06 d $<$ $P_{\rm orb}$ $\leq$ 0.07 d.  Open 
circles and rectangles represent our results in ASASSN-15jd 
and the amplitudes of the candidates for a period bouncer, 
respectively.  
Other data indicated by filled circles are the same as in Fig.~87 
of \citet{Pdot3}.  
Right: Dependence of superhump amplitudes on orbital periods (the 
relation between the maximum superhump amplitudes and the orbital 
period).  Open circles and rectangles represent the same as in 
the left panel.  Filled circles indicate the amplitudes of ordinary 
SU UMa-type systems got from Fig.~89 of \citet{Pdot3}.   
We selected epochs 5 $<$ $E$ $<$ 10 to illustrate the maximum superhump 
amplitudes.  The curve indicates a spline-smoothed interpolation 
of the data of ordinary SU UMa-type stars.}
\label{amplitude}
\end{figure*}

   We show the variation of superhump amplitudes of the SU UMa-type systems 
whose orbital periods are 0.06--0.07 d including ASASSN-15jd and the 
candidates for a period bouncer, in the left panel of Figure 
\ref{amplitude}.  
In plotting this figure, we measured the amplitudes using the template 
fitting method described in \citet{Pdot} and took zero of the cycle
count from the start of stage B.  In ASASSN-15jd, we regarded that the 
maximum amplitude on BJD 2457167 is at $E=0$ for comparison since we 
could not clearly identify the start of stage B.  Previous research 
demonstrates that the mean superhump amplitude is small 
($\lesssim$0.1 mag) in the extreme WZ Sge-type systems which are the 
candidates for a period bouncer \textcolor{black}{(see Table 1 for the details)}.  
From this figure, we can see the amplitudes in ASASSN-15jd are about two 
times smaller than those of ordinary SU UMa-type DNe and have about the
same values as those in the best candidates for a period bouncer.  

   Additionally, in the right panel of Figure \ref{amplitude}, we give 
the relation between maximum superhump amplitudes and orbital periods in 
SU UMa-type stars.  The samples of ordinary SU UMa-type objects and 
period-bouncer candidates that we used in this figure are the objects 
listed in Table 73 of \citet{Pdot3} and \textcolor{black}{Table 1 of this 
paper, respectively}.  
In this figure, we substituted the stage B superhump period and the average 
superhump period for the orbital periods in OT J060009.9$+$142615 and 
ASASSN-15jd, respectively.  
The mean maximum amplitude in the objects which have $P_{\rm orb}$ = 
0.065 is about 0.25 mag, but, that in ASASSN-15jd is 0.091.  
In addition, the maximum amplitudes in the best candidates for a 
period bouncer are also small and out of the distribution of ordinary
SU UMa-type DNe.  

Therefore, the small superhump amplitude in ASASSN-15jd can be 
regarded as the representation of the characteristic properties of a 
period bouncer.  
   The properties of the candidates for a period bouncer are 
summarized in Table 1.  

\subsection{Absence of Early Superhumps}

   In ASASSN-15jd, no ordinary superhumps were observed for about ten days 
from the onset of the outburst to the small dip of brightness (on BJD 
2457165).  In addition, we could not detect early superhumps for that term although 
they are usually observed in outbursts of WZ Sge-type stars for about 
a week \citep{kat15wzsge} before ordinary superhumps develop.  
   \citet{uem12ESHrecon} proposed that early superhumps are caused 
by the rotation effect of non-axisymmetrically flaring accretion disks 
(see also \cite{nog97alcom} and \cite{kat02wzsgeESH}).  
High-inclination (edge-on) systems thus tend to show large-amplitude 
early superhumps.  
   In addition, early superhumps are regarded as the manifestation 
of the 2:1 resonance tidal instability and the 2:1 resonance is 
considered to suppress the development of ordinary superhumps
\citep{lub91SHa,osa02wzsgehump,osa03DNoutburst}.  
Based on the above theories, we suggest that the inclination angle of 
ASASSN-15jd is low and we observed this system at nearly face-on from 
our results: 1) the superhumps start to develop at the dip, 2) neither 
early superhumps nor ordinary ones were observed before the dip  

   Note that the duration for which the 2:1 resonance supposed to have 
been dominant before the superhumps developed in ASASSN-15jd is longer 
than the average duration of early superhumps in ordinary WZ Sge-type DNe.  
In the superoutbursts of the promising candidates for a period bouncer, 
the durations of the early superhumps were usually long and this can 
be one of characters of period bouncers 
(\cite{kat13j1222,nak13j2112j2037}; Nakata et al.~in preparation).  
Therefore, this finding suggests that ASASSN-15jd has yet another property 
common to period bouncers.

\section{Summary}

   We have reported on our photometric observations of a WZ Sge-type 
DN, ASASSN-15jd.  What we found on this object from our observations 
is as follows: 
\begin{itemize}
\item ASASSN-15jd is the first WZ Sge-type object with a small dip of 
brightness in the middle of the superoutburst.  This implies that the 3:1 
resonance grew up slowly in its outburst and the mass ratio is considered 
to be small.  
\item The superhump amplitude in ASASSN-15jd was small and the mean 
value was less than 0.1 mag.  
\item The inclination angle of ASASSN-15jd may be low judging 
from the lack of early superhumps.  The interval without ordinary 
superhumps was relatively long, about ten days.  
\end{itemize}
   We suggest that ASASSN-15jd would be a promising candidate for a 
period bouncer because the above properties are very similar to 
those of other best candidates for a period bouncer.  
\citet{gan09SDSSCVs} reported that the optical spectra of SDSS 
CVs whose orbital periods are close to the minimum one are dominated 
by emission from the white dwarf photosphere and that the contribution 
to the spectra from the companion star is no or little.  This means that 
they have very low mass companion stars, in other words, they have very 
small mass ratios.  
Hence, spectroscopic observation of ASASSN-15jd may be useful to confirm 
its low mass ratio.

\section*{Acknowledgements}

This work was supported by a Grant-in-Aid ``Initiative for 
High-Dimensional Data-Driven Science through Deepening of 
Sparse Modeling'' from the Ministry of Education, Culture, 
Sports, Science and Technology (MEXT) of Japan (25120007). 
Also, it was partially supported by the RFBR grant 15-02-06178.  
We appreciate All-Sky Automated Survey for Supernovae (ASAS-SN) 
detecting a large amount of DNe.  
We are thankful to many amateur observers for providing a lot of 
data used in this research.

\section*{Supporting information}

Additional supporting information can be found in the online version 
of this article:
Supplementary tables 2, 3 and 4.

Supplementary data is available at PASJ Journal online 
({\it included at the end in this astro-ph version}).

\newcommand{\noop}[1]{}

\begin{table*}[b]
  \caption{List of Instruments}
\label{tab:listtel}
\begin{center}
\begin{tabular}{lccc}
\hline
CODE$^{*}$ & Telescope (\& CCD) & Observatory (or Observer) & Site\\
\hline
CRI & 38cm K-380+Apogee E47 & Crimean astrophysical observatory & Crimea \\
DPV & 28cmSC+MII G2-1600 & Astronomical Obs. on Kolonica Saddle & Slovakia \\
    & 35cmSC+MII G2-1600 & Astronomical Obs. on Kolonica Saddle & Slovakia \\
    & VNT 1m+FLI PL1001E & Astronomical Obs. on Kolonica Saddle & Slovakia \\
Ter & Zeiss-600 60cm & Terskol Observatory & Russia \\
    & S2C 35cm      & Terskol Observatory & Russia \\
RPc & FTN 2.0m+E2V 42-40 & LCOGT$^{\dagger}$ & Hawaii, USA \\
    & 35cmSC+SXV-H9 CCD & Roger D. Pickard & UK \\
Trt & 25cm ALCCD5.2 (QHY6) & Tam\'as Tordai & Budapest, Hungary \\
KU1 & 40cmSC+ST-9XEI & Kyoto U. Team & Kyoto, Japan  \\
SWI & C14 35cmSC+ST10XME & William L. Stein & New Mexico, USA \\
OKU & 51cm+Andor DW936N-BV & OKU Astronomical Observatory & Osaka, Japan \\
IMi & 35cmSC+SXVR-H16 & Ian Miller & Furzehill Observatory \\
Van & 35cmSC+SBIG ST-7XME CCD & CBA Belgium Observatory & Belgium \\
DKS & LX200 25cm & Rolling Hills Observatory & USA \\
Ioh & 30cmSC+ST-9XE CCD & Hiroshi Itoh & Tokyo, Japan \\
\hline
\multicolumn{4}{l}{
$^{*}$see the annotation in Table 3} \\
\multicolumn{4}{l}{
$^{\dagger}$Las Cumbres Observatory Global Telescope Network} \\
\end{tabular}
\end{center}
\end{table*}

\begin{table*}[b]
\caption{Log of observations of ASASSN-15jd in 2015}
\label{log}
\begin{center}
\begin{tabular}{rrrrrccr}
\hline
${\rm Start}^{*}$ & ${\rm End}^{*}$ & ${\rm Mag}^{\dagger}$ & ${\rm Error}^{\ddagger}$ & $N^{\S}$ & ${\rm Obs}^{\parallel}$ & ${\rm Band}^{\#}$ & exp[s] \\ \hline
58.3835 & 58.5979 & 14.300 & 0.002 & 173 & Van & $C$ & 60\\
59.4360 & 59.5155 & 14.386 & 0.002 & 93 & Van & $C$ & 60\\
60.3841 & 60.5916 & 14.484 & 0.001 & 227 & Van & $C$ & 60\\
61.3155 & 61.5612 & 14.523 & 0.001 & 306 & DPV & $C$ & 60\\
61.3888 & 61.5035 & 14.517 & 0.001 & 153 & Trt & $C$ & 60\\
61.5031 & 61.5974 & 14.540 & 0.001 & 95 & Van & $C$ & 60\\
62.0047 & 62.1933 & 14.530 & 0.001 & 492 & KU1 & $C$ & 30\\
62.1640 & 62.1899 & 14.500 & 0.001 & 70 & OKU & $C$ & 30\\
62.4018 & 62.5906 & 14.571 & 0.001 & 192 & Van & $C$ & 60\\
62.4563 & 62.5377 & 14.576 & 0.001 & 150 & IMi & $CV$ & 30\\
63.2956 & 63.4021 & 14.674 & 0.002 & 100 & CRI & $C$ & 120\\
63.3849 & 63.5911 & 14.707 & 0.001 & 231 & Van & $C$ & 60\\
64.1458 & 64.2850 & 14.706 & 0.002 & 275 & KU1 & $C$ & 30\\
64.1566 & 64.2693 & 14.732 & 0.001 & 156 & OKU & $C$ & 30\\
65.2691 & 65.3961 & 15.423 & 0.009 & 61 & CRI & $C$ & 120\\
65.3866 & 65.4792 & 15.476 & 0.003 & 107 & Van & $C$ & 60\\
66.3356 & 66.5482 & 14.906 & 0.006 & 145 & CRI & $C$ & 120\\
66.3865 & 66.5916 & 14.844 & 0.003 & 195 & Van & $C$ & 60\\
66.4446 & 66.5118 & 14.843 & 0.002 & 126 & IMi & $CV$ & 30\\
66.5903 & 66.8343 & 14.787 & 0.003 & 268 & DKS & $C$ & 60\\
67.1612 & 67.2776 & 14.661 & 0.004 & 149 & OKU & $C$ & 30\\
67.2943 & 67.5335 & 14.655 & 0.004 & 110 & CRI & $C$ & 120\\
67.3753 & 67.5206 & 14.643 & 0.002 & 359 & Ter & $C$ & 30\\
67.9861 & 68.2617 & 14.690 & 0.001 & 719 & KU1 & $C$ & 30\\
68.1661 & 68.2851 & 14.695 & 0.002 & 318 & OKU & $C$ & 30\\
68.2725 & 68.3749 & 14.696 & 0.003 & 73 & CRI & $C$ & 120\\
68.6699 & 68.8851 & 14.700 & 0.002 & 272 & SWI & $C$ & 60\\
69.0767 & 69.2842 & 14.960 & 0.002 & 288 & Ioh & $C$ & 60\\
69.1712 & 69.2695 & 14.722 & 0.001 & 246 & OKU & $C$ & 30\\
69.4363 & 69.5539 & 14.777 & 0.002 & 228 & IMi & $CV$ & 30\\
69.5811 & 69.6179 & 14.812 & 0.006 & 52 & RPc & $V$ & 60\\
70.1466 & 70.2738 & 14.856 & 0.002 & 336 & OKU & $C$ & 30\\
70.6038 & 70.8384 & 14.938 & 0.002 & 279 & DKS & $C$ & 60\\
71.3427 & 71.4869 & 15.007 & 0.002 & 186 & DPV & $C$ & 60\\
71.3710 & 71.4926 & 15.002 & 0.003 & 169 & Trt & $C$ & 60\\
71.5756 & 71.7965 & 15.039 & 0.002 & 256 & DKS & $C$ & 60\\
72.2365 & 72.3010 & 15.061 & 0.005 & 141 & KU1 & $C$ & 30\\
74.4216 & 74.5255 & 16.666 & 0.010 & 114 & RPc & $CV$ & 60\\
75.5129 & 75.5338 & 18.999 & 0.159 & 28 & DPV & $C$ & 60\\
77.1767 & 77.1926 & 18.298 & 0.036 & 9 & OKU & $C$ & 30\\
86.1662 & 86.2473 & 15.812 & 0.019 & 40 & OKU & $C$ & 30\\
86.4214 & 86.5462 & 15.364 & 0.001 & 128 & Van & $C$ & 60\\
86.6490 & 86.8169 & 15.360 & 0.002 & 215 & SWI & $C$ & 120\\
87.4142 & 87.5649 & 16.265 & 0.006 & 128 & Van & $C$ & 60\\
\hline
\multicolumn{8}{l}{$^{*}{\rm BJD}-2457100.0$.}\\
\multicolumn{8}{l}{$^{\dagger}$Mean magnitude.}\\
\multicolumn{8}{l}{$^{\ddagger}1\sigma$ of mean magnitude.}\\
\multicolumn{8}{l}{$^{\S}$Number of observations.}\\
\multicolumn{8}{l}{$^{\parallel}$Observer's code: Van (Tonny Vanmunster), DPV (Pavol A. Dubovsky),}\\
\multicolumn{8}{l}{Trt (Tam\'as Tordai), KU1 (Kyoto Univ. Team),}\\
\multicolumn{8}{l}{OKU (Osaka Kyoiku Univ. Team), IMi (Ian Miller),}\\
\multicolumn{8}{l}{CRI (Crimean Observatory Team), DKS (Shawn Dvorak)}\\
\multicolumn{8}{l}{Ter (Terskol Observatory), SWI (William L. Stein) Ioh (Hiroshi Itoh),}\\
\multicolumn{8}{l}{and RPc (Roger D. Pickard)}\\
\multicolumn{8}{l}{$^{\#}$Filter. ``$V$'' means $V$ filter,``$C$'' and $CV$ mean no filter (clear).}\\
\end{tabular}
\end{center}
\end{table*}

\begin{table*}[b]
\caption{Times of superhump maxima in ASASSN-15jd}
\label{max}
\begin{center}
\begin{tabular}{rllrr}
\hline
$E$ & ${\rm Max}^{\dagger}$ & Error & ${O - C}^{\ddagger}$ & $N^{\S}$  \\ \hline
0 & 57166.4601 & 0.0008 & 0.0009 & 136 \\
1 & 57166.5191 & 0.0007 & -0.0051 & 84 \\
2 & 57166.5840 & 0.0011 & -0.0051 & 56 \\
3 & 57166.6536 & 0.0011 & -0.0006 & 64 \\
4 & 57166.7186 & 0.0007 & -0.0006 & 60 \\
5 & 57166.7833 & 0.0004 & -0.0010 & 63 \\
11 & 57167.1774 & 0.0004 & 0.0030 & 59 \\
12 & 57167.2401 & 0.0003 & 0.0007 & 61 \\
13 & 57167.3122 & 0.0009 & 0.0077 & 21 \\
14 & 57167.3727 & 0.0005 & 0.0033 & 87 \\
15 & 57167.4349 & 0.0003 & 0.0004 & 154 \\
16 & 57167.4968 & 0.0003 & -0.0027 & 136 \\
24 & 57168.0244 & 0.0005 & 0.0047 & 137 \\
25 & 57168.0914 & 0.0007 & 0.0066 & 131 \\
26 & 57168.1518 & 0.0004 & 0.0020 & 169 \\
27 & 57168.2164 & 0.0003 & 0.0016 & 274 \\
28 & 57168.2821 & 0.0007 & 0.0022 & 121 \\
19 & 57168.3470 & 0.0011 & 0.0021 & 36 \\
35 & 57168.7369 & 0.0004 & 0.0019 & 63 \\
36 & 57168.7997 & 0.0004 & -0.0003 & 68 \\
41 & 57169.1251 & 0.0009 & -0.0001 & 72 \\
42 & 57169.1894 & 0.0004 & -0.0008 & 177 \\
43 & 57169.2517 & 0.0006 & -0.0035 & 189 \\
46 & 57169.4495 & 0.0004 & -0.0008 & 77 \\
47 & 57169.5165 & 0.0005 & 0.0012 & 99 \\
57 & 57170.1691 & 0.0004 & 0.0035 & 126 \\
58 & 57170.2299 & 0.0003 & -0.0007 & 139 \\
64 & 57170.6177 & 0.0016 & -0.0030 & 44 \\
65 & 57170.6834 & 0.0017 & -0.0024 & 64 \\
76 & 57171.3983 & 0.0005 & -0.0028 & 140 \\
77 & 57171.4634 & 0.0010 & -0.0027 & 133 \\
79 & 57171.5994 & 0.0019 & 0.0033 & 60 \\
80 & 57171.6622 & 0.0009 & 0.0010 & 65 \\
81 & 57171.7251 & 0.0011 & -0.0011 & 56 \\
82 & 57171.7902 & 0.0016 & -0.0011 & 38 \\
89 & 57172.2470 & 0.0011 & 0.0006 & 79 \\
\hline
\multicolumn{5}{l}{$^{*}$Cycle counts.}\\
\multicolumn{5}{l}{$^{\dagger}$BJD$-$2400000.0.}\\
\multicolumn{5}{l}{$^{\ddagger}$ C = 2457166.459117$+$0.0650258 E.}\\
\multicolumn{5}{l}{$^{\S}$Number of points used for determining}\\
\multicolumn{5}{l}{the maximum.}\\
\end{tabular}
\end{center}
\end{table*}

\end{document}